\documentclass[aps,prl,
longbibliography,
reprint
]{revtex4-2}
\pdfoutput=1

\usepackage{amsfonts}
\usepackage{amsmath}
\usepackage{amssymb}
\usepackage{amsthm}
\usepackage[mathscr]{eucal}
\usepackage{bbold}
\usepackage{braket}
\usepackage{color}
\usepackage{dsfont}
\usepackage{framed}

\usepackage{mathtools}
\usepackage{physics}
\usepackage[normalem]{ulem}
\usepackage{stackrel}
\usepackage{tensor}
\usepackage{tikz-cd}
\usepackage{ulem}

\usepackage{hyperref}

\definecolor{SHCcolor}{rgb}{0.9,0.1,0.1}

\definecolor{GHcolor}{rgb}{0.2,0,0.9}

\definecolor{GTcolor}{rgb}{0.2,0.9,0.2}

\definecolor{DWcolor}{rgb}{0.7,0.2,0.7}

\newcommand{\ignore}[1]{}

\newcommand{\evalat}[2]{\left. #1 \right|_{#2}}

\newcommand{\im}{\mathrm{i}}

\makeatletter\def\@fpheader{~}\makeatother
\newcommand{\Zbulk}{\mathcal{Z}_{\tiny \text{bulk}}}
\newcommand{\Zbdy}{\mathcal{Z}_{\tiny \text{bdy}}}

\newcommand{\PIbdy}[1]{\expval{#1}_{\tiny \text{bdy}}}
\newcommand{\PIg}[1]{\expval{#1}_{g}}

\hypersetup{
	colorlinks,
	linkcolor={blue!80!black},
	citecolor={blue!80!black},
	urlcolor={blue!80!black}
}

\begin{document}

	\preprint{arXiv:2103.05014 [hep-th]}
	
	\title{Boundary Causality Violating Metrics in Holography}
	\author{Sergio~Hern\'andez-Cuenca, Gary~T.~Horowitz, Gabriel~Trevi\~no, and Diandian~Wang}
	\affiliation{Department of Physics, University of California, Santa Barbara, CA 93106}

	\begin{abstract}
		Even for holographic theories that obey boundary causality, the full bulk Lorentzian path integral includes metrics that violate this condition. This leads to the following puzzle: The commutator of two field theory operators at spacelike-separated points on the boundary must vanish. However, if these points are causally related in a bulk metric, then the bulk calculation of the commutator will be nonzero. It would appear that the integral over all metrics of this commutator must vanish exactly for holography to hold. This is puzzling since it must also be true if the commutator is multiplied by any other operator. Upon a careful treatment of boundary conditions in holography, we show how the bulk path integral leads to a natural resolution of this puzzle.
	\end{abstract}
	
	\maketitle

	\section{Introduction}
	
	A bulk description of nonperturbative quantum gravity is not yet available. A popular approach is to consider a path integral over metrics. We investigate some aspects of this approach. The metric could turn out to be only a low-energy approximation of other fundamental degrees of freedom. Our results would plausibly still apply, since the issues we address arise already in metrics with curvature well below the Planck (and string) scale.
	
	We begin with a Lorentzian formulation of holography, in which one integrates over asymptotically anti-de Sitter (AdS) spacetimes and matter fields (with certain boundary conditions) to compute correlation functions in a dual quantum field theory. One often works in a large $N$ or semiclassical limit and only includes classical supergravity solutions and small perturbations of them. Classically, spacetimes satisfy the null energy condition so the Gao-Wald theorem \cite{Gao:2000ga} ensures that the bulk preserves boundary causality. In other words, the fastest way to send a signal between two observers on the boundary is via a path that stays on the boundary. No trajectory that enters the bulk can arrive sooner. Semiclassically, the achronal averaged null energy condition ensures boundary causality \cite{Akers:2016ugt}. (A necessary and sufficient geometric condition is given in \cite{Engelhardt:2016aoo}.)
	
	However, in full quantum gravity (finite $N$), the bulk path integral includes metrics that violate boundary causality. That is, two boundary points that are spacelike-separated on the boundary can nevertheless be timelike-separated with respect to some bulk metrics. We will see that these boundary causality violating metrics are not ``rare'' -- they include open sets in the space of metrics. These metrics describe causally well-behaved bulk geometries -- there are no closed timelike curves or any causal pathology. So there is no reason to exclude them from the bulk path integral. Hence one might worry that they could contribute to causality violations in the dual field theory, which would be a problem.
	
	Consider, for example, the commutator of an operator $\mathcal{O}$ at two spacelike-separated boundary points: $ [\mathcal{O}(x),\mathcal{O}(y)]$. This must vanish identically, but holography says that it should be given by a bulk path integral over metrics and a matter field $\phi$ dual to $\mathcal{O}$. If one computes the commutator $[\phi(p), \phi(q)]$ in each metric $g$, and then follows the standard limiting procedure of taking $p \to x$ and $q\to y$, one generically expects a nonzero answer whenever $x$ and $y$ are causally related with respect to $g$. This is not yet a contradiction: the integral of this over metrics $g$ weighted by $e^{\im S_g}$ could still vanish exactly.
	
	However, it is not just one quantity that must integrate to zero. In principle, $\mathcal{O}$ can be any operator in the dual field theory, and the commutator could be multiplied by any other operator. Their corresponding bulk expressions would all be nonzero in some metrics $g$, but their integrals over $g$ must still vanish identically. Since the gravitational weighting $e^{\im S_g}$ and measure are independent of the operator insertions, this seems unlikely.
	
	By studying the boundary conditions for the bulk path integral computation of commutators, we identify some freedom whenever the boundary theory is causal and unitary. We show that there is a way to use this freedom to make  boundary causality  manifest.
	
	
	\section{Specific examples}
	
	For $b\in\mathbb{R}$, consider the family
	\begin{equation}
		\label{eq:fmetric}
		g_{(b)} = - f_b(r) \,\dd t^{2} + f_b(r)^{-1} \,\dd r^{2} + r^{2} \,\dd\Omega_2^{2}.
	\end{equation}
	One requires $f_b(r)\approx r^2/\ell^2$ asymptotically, $f_b(0)=1$ and $f_b'(0)=0$ for differentiability at $r=0$.
	We connect this family at $b=0$ to AdS$_4$ by setting $f_0(r) =  r^2/\ell^2+1$.
	In pure AdS, two boundary points are null-separated on the boundary if and only if they are null-separated in the bulk. Hence we explore boundary causality in \eqref{eq:fmetric} by comparing to AdS. Restricting to radial null geodesics, the boundary-to-boundary crossing time is
	\begin{equation}
		\label{eq:crosstime}
		t_{\infty}=2\int_{0}^{\infty} \frac{\dd r}{f_b(r)}.
	\end{equation}
	Hence \eqref{eq:fmetric} violates boundary causality if $t_{\infty}<t_{\infty}^{\text{AdS}}=\pi\ell$; e.g., $f_b(r)>f_0(r)$ everywhere would suffice. For instance, setting $\ell=1$, \eqref{eq:fmetric} with $f_b(r)=r^2 +1+b \; r^{2}(1+r^{4})^{-1}$ violates boundary causality for $b>0$.
	
	These static configurations only contribute to a path integral where the initial and final surfaces have induced metric $\dd r^2 /f_b(r) + r^2 \dd\Omega_2^2$. One might ask if there exist classical solutions interpolating between such surfaces. The answer is yes: global AdS gives the desired induced metric with $f_b(r)\ge f_0(r)$ on spacelike surfaces with $t=b\, h(r)$ where $h$ must just fall off like $1/r^2$ or faster.
	
	Alternatively, to match a static AdS slice, one can just make $b$ time-dependent near the surface, going to zero on it. More generally, one can match any given induced metric by making $g_{(b)}$ appropriately time-dependent without affecting the causality violating region.
	
	Although we have focused on four-dimensional examples, it is clear that similar examples exist in all dimensions greater than two. In particular, if $x$ is timelike-related to $y$ in one metric, it remains timelike-related in any nearby metric. Hence boundary causality is violated in open sets in the space of metrics.


	\section{Operator dictionaries}
	
	Holographic duality is generally formulated as an equivalence between bulk and boundary partition functions \cite{Gubser:1998bc,Witten:1998qj}:
	\begin{equation}\label{eq:duality}
		\Zbulk[\phi_0] = \Zbdy[\phi_0].
	\end{equation}
	Here, $\Zbdy$ is defined on some fixed Lorentzian manifold $(\mathcal{B},\gamma)$, whereas $\Zbulk$ path-integrates over spacetimes $(\mathcal{M},g)$ having $( \mathcal{B},\gamma)$ as their conformal boundary. The symbol $\phi_0$ is a placeholder for all boundary conditions for dynamical fields on the bulk side, and for sources of operator deformations of the action on the boundary side. To make progress, most work in the literature studies \eqref{eq:duality} perturbatively in $1/N$, where $\Zbulk$ is amenable to a saddlepoint approximation (plus corrections). Here we test one aspect of \eqref{eq:duality} as a statement of holography in full quantum gravity, i.e., nonperturbatively in $1/N$. 
	
	To study real-time correlation functions, we assume \eqref{eq:duality} applies in Lorentzian signature. From the standpoint of the boundary theory, $\Zbdy[\phi_0]$ is a standard field-theoretic generating functional. Explicitly \cite{Witten:1998qj},
	\begin{equation}
		\label{eq:zbdy}
		\Zbdy[\phi_0] = \PIbdy{\exp{\im\int_{\mathcal{B}} \; \sum_\alpha \,\phi_0^\alpha \, \mathcal{O}_\alpha}},
	\end{equation}
	where $\PIbdy{\,\cdot\,}$ denotes the boundary path integral over quantum fields, and each $\mathcal{O}_\alpha$ is a gauge-invariant operator sourced by $\phi_0^\alpha$. In light of \eqref{eq:duality}, the bulk object $\Zbulk[\phi_0]$ may also be regarded as a generating functional of boundary correlators of the form
	\begin{equation}
		G\equiv\PIbdy{\mathcal{O}_n(x_n) \cdots \mathcal{O}_1(x_1)}.
	\end{equation}
	If $G$ is time-ordered with respect to boundary time, then
	\begin{equation}\label{eq:todiff}
		G = \evalat{\frac{1}{\im^n} \frac{\delta}{\delta \phi^n_0(x_n)} \cdots \frac{\delta}{\delta \phi^1_0(x_1)} \; \frac{\Zbulk[\phi_0]}{\Zbulk[0]}}{\phi_0=0}.
	\end{equation}
	The rationale so far is that of \cite{Gubser:1998bc,Witten:1998qj}, later coined the ``differentiate dictionary'' in \cite{Harlow:2011ke}.
	
	In practice though, one usually takes a leap and follows the logic of \cite{Susskind:1998dq,Banks:1998dd} to argue that a boundary correlator should be nothing but an appropriate limit of bulk ones. More specifically, if $\phi_\alpha$ is the bulk field dual to $\mathcal{O_\alpha}$ and the latter has conformal dimension $\Delta_\alpha$, one may propose~\footnote{Time-ordering has been intentionally omitted below since the notion of time order in the bulk is subtle -- see the next section.}
	\begin{equation}\label{eq:toextra1}
		G \stackrel{?}{\propto}
		\int \mathcal{D} g \, e^{\im S_g} \lim_{z\to0} z^{\Delta_\Sigma} \PIg{\phi_n(x_n,z)\cdots\phi_1(x_1,z)},
	\end{equation}
	where $\Delta_\Sigma = \sum_\alpha\Delta_\alpha$, $\PIg{\,\cdot\,}$ denotes the bulk path integral over quantum fields in a fixed metric $g$, and $(x,z)$ are Fefferman-Graham coordinates \cite{fefferman_1985}. (These provide an unambiguous asymptotic location for the bulk operators.)
	Alternatively, one could consider \eqref{eq:toextra1} with $z\to0$ taken outside the integral.
	While physically both seem like reasonable proxies for boundary correlators, it is unclear whether they are equal or mathematically consistent with \eqref{eq:todiff}. 
	
	When gravity is treated semiclassically, these two options do agree. In this case, they were referred to as the ``extrapolate dictionary'' in \cite{Harlow:2011ke}, where their consistency with \eqref{eq:todiff} was carefully studied. We deviate from \cite{Harlow:2011ke} by studying Lorentzian physics and at a nonperturbative level.
	
	For any causal field theory, local operators at spacelike boundary separation commute. In the bulk, there appears to be a problem with the extrapolate dictionary stemming from the presence of off-shell contributions to the gravitational path integral that violate boundary causality.
	Similar problems arise regardless of when one takes the $z\to0$ limit, so consider \eqref{eq:toextra1} for definiteness. For a commutator of operators at spacelike separation, $\PIbdy{ [\mathcal{O}(x), \mathcal{O}(y)] }=0$, so \eqref{eq:toextra1} becomes
	\begin{equation}\label{eq:fine0}
		\int \mathcal{D} g\, e^{\im S_g} \lim_{z\to0} z^{2\Delta} \PIg{[\phi(x,z),\phi(y,z)]} \stackrel{?}{=} 0.
	\end{equation}
	As explained above, there are open sets of bulk metrics for which $x$ and $y$ are causally related and the commutator is generically nonzero. So why should the integral vanish exactly?

	
	\section{Resolution}
	\label{sec:resol}
\begin{figure}[t]
		\centering
		\includegraphics[width=0.483\textwidth]{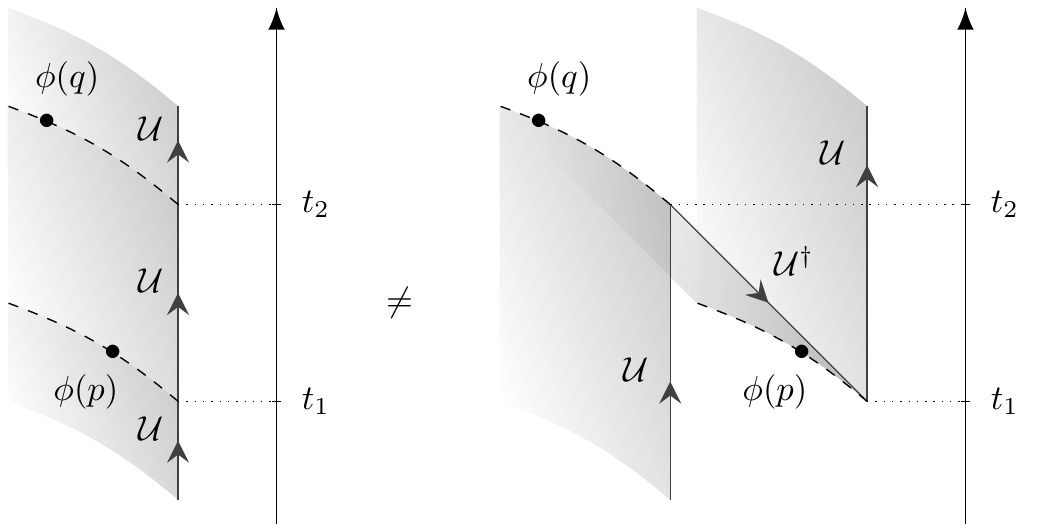}
		\caption{Inequivalent timefolds for computing a two-point function for a field $\phi$ inserted with different time orderings. Point $q$ is to the future of $p$, each respectively lying at times $t_1$ and $t_2>t_1$. Left: time-ordered correlator $\expval{\phi(q)\phi(p)}$, where time-evolution is implemented by $\mathcal{U}$ everywhere. Right: nontrivial timefold for the out-of-time-order correlator $\expval{\phi(p)\phi(q)}$, with backwards time-evolution via $\mathcal{U}^\dagger$.}
		\label{fig:foldillus}
	\end{figure}

	We now explain an important subtlety omitted from the above discussion.
	Consider the commutator
	\begin{equation}
		\label{eq:comm}
		\PIg{[\phi(p),\phi(q)]} \equiv \PIg{\phi(p)\phi(q)}-\PIg{\phi(q)\phi(p)},
	\end{equation}
	with each term computed via a path integral. Although we have used the bulk notation $\PIg{\,\cdot\,}$, this also applies to the boundary by replacing $(\mathcal{M},g)$ with $\mathcal{(B,\gamma)}$. Now, the path integral on the original manifold always computes time-ordered correlators.
	To compute an out-of-time-order correlator, one needs to manufacture from the original manifold a \textit{timefold} as follows. Consider an $n$-point function for a field $\phi$,
	\begin{equation}
		\PIg{\phi_{+}|\phi(p_n)\cdots\phi(p_{2})\phi(p_1)|\phi_{-}},
	\end{equation}
	where $\phi_\pm$ are initial and final states on Cauchy surfaces $\Sigma_\pm$. Starting at $\Sigma_-$, one should use the evolution operator $\mathcal{U}$ up to a Cauchy surface containing $p_1$ and insert $\phi(p_1)$. One should then evolve to $p_2$ for the next insertion. If $p_2$ is to the future of $p_1$, one could just use $\mathcal{U}$ again. However, if $p_2$ is to the past of $p_1$, one should evolve backwards with $\mathcal{U}^\dagger$ for the insertion of $\phi(p_2)$. This procedure must be repeated for all $n$ insertions before finally evolving to $\Sigma_+$. The whole process produces a timefold, a new zigzagged spacetime implementing the correct ordering of operators (see Fig.\;\ref{fig:foldillus}).
	In the path integral, backward components receive weighting $e^{-\im S_g}$, instead of the usual $e^{\im S_g}$ on forward ones. 
	
	If two points are timelike-related, the commutator is a difference between two correlators computed on distinct timefolds, so will be generically nonzero. In the case of spacelike-separated points, however, the two points can be inserted on the same Cauchy surface. Hence the same timefold qualifies for a computation of either ordering, yielding a vanishing commutator. Again, the discussion here applies to both boundary commutators $\PIbdy{[\mathcal{O}(x),\mathcal{O}(y)]}$ and bulk commutators $\PIg{[\phi(x,z),\phi(y,z)]}$ on a specific metric, and therefore one has both boundary and bulk timefolds.
	
	In general, the field theory $\Zbdy$ is naturally defined on some manifold $\mathcal{B}$, and $\Zbdy[\mathcal{B},\phi_0]$ is the generating functional of time-ordered correlators on $\mathcal{B}$. To compute an out-of-time-order correlator, one needs to extend $\Zbdy$ to an appropriate timefold spacetime $\mathcal{F}$ constructed out of $\mathcal{B}$ as described above. Then $\Zbdy[\mathcal{F},\phi_0]$ allows one to compute the desired correlator. The statement of holography in terms of partition functions must account for this. Hence one has to more explicitly rewrite \eqref{eq:duality} as 
	\begin{equation}\label{eq:equalF}
		\Zbulk[\mathcal{F},\phi_0] = \Zbdy[\mathcal{F},\phi_0],
	\end{equation}
	where $\Zbulk[\mathcal{F},\phi_0]$ path-integrates over all metrics with the timefold $\mathcal{F}$ as conformal boundary, and  boundary conditions $\phi_0$ appropriately distributed on $\mathcal{F}$.
	
	We will call the regions of spacetime separated by the creases the ``sheets'' of the timefold. For a causal and unitary boundary theory, if $x$ and $y$ are spacelike-separated, $\expval{ \mathcal{O}(y) \mathcal{O}(x)}$ can be computed on either a trivial timefold (i.e. the original spacetime with no foldings), or a nontrivial timefold with $\mathcal{O}(x)$ on some sheet and $\mathcal{O}(y)$ on some other sheet with the same result. For example, evolution forward and back -- without operator insertions -- is the identity.  It is not obvious that the corresponding bulk partition functions will agree since we do not know (independent of holography) that the gravitational path integral describes unitary evolution. This leads to a potential ambiguity in the definition of $\Zbulk$. Since we are trying to understand how a basic property of the boundary theory follows from the bulk path integral, we have to resolve this ambiguity. If the boundary theory is causal and unitary, we adopt a \textit{minimal-timefold} approach, introducing a fold only when needed to represent operators at causally-related points where the later operator appears first in the correlator. When folding minimally, backward evolution just needs to sweep causal diamonds between insertions at non-time-ordered points (including small neighborhoods for spacelike creases). Minimal timefolds are introduced to disambiguate inequivalent computations which should nonetheless give equivalent results -- if the boundary theory is acausal or non-unitary, however, results will generically disagree and the minimal-timefold prescription does not apply.

	We now show that with the minimal timefold, both the differentiate and extrapolate dictionaries predict no violation of boundary causality.
	Consider first the differentiate dictionary. Equation \eqref{eq:todiff} tells us how the time-ordered correlators on the boundary are computed from the bulk, but we have so far not included out-of-time-order correlators in this equality. Extending the dictionary to general $n$-point functions, one has
	\begin{equation}\label{eq:diffcon}
		G = \evalat{\frac{1}{\im^n} \frac{\delta}{\delta \phi^n_0(x_n)} \cdots \frac{\delta}{\delta \phi^1_0(x_1)} \; \frac{\Zbulk[\mathcal{F},\phi_0]}{\Zbulk[\mathcal{F},0]}}{\phi_0=0},
	\end{equation}
	where $\mathcal{F}$ is a minimal \textit{boundary} timefold constructed so as to order the operator insertions as given in $G$ (cf. \eqref{eq:todiff} when the left-hand side is time-ordered, in which case $\mathcal{F}$ is just $\mathcal{B}$).
	If $x$ and $y$ are spacelike-separated boundary points, the minimal timefold for a two-point correlator is trivial. Since this is the same for both terms in the commutator, the only difference between them is the order in which one takes the derivatives with respect to $\phi_0$. Since these derivatives commute, the commutator obviously vanishes. Note that this is completely independent of whether $x$ and $y$ are causally related with respect to some bulk metrics. All we use is that the integral over all metrics and matter fields is some functional of the boundary conditions $\phi_0$ on a trivial timefold. If $x$ and $y$ are causally related on the boundary, the situation is different. In that case, each term in the commutator requires a different minimal timefold: a trivial one for the time-ordered correlator, and a nontrivial one for the out-of-time-order correlator. In the latter, the source for the earlier operator is moved to the second sheet, so the commutator can be nonzero.

	We now turn to the extrapolate dictionary as in \eqref{eq:toextra1}.
	For a bulk metric relating $x$ and $y$ causally, one would expect $\PIg{\phi(x,z)\phi(y,z)}\ne \PIg{\phi(y,z)\phi(x,z)}$ since they are computed on different bulk timefolds. This nonzero commutator on $g$ is the origin of the causality puzzle. However, there is a subtlety: The boundary condition on $g$ must be the same as for the differentiate dictionary. Namely, one must have a trivial boundary timefold when $x$ and $y$ are spacelike-related on the boundary. If one computes the commutator by a path integral over fields on the bulk metric $g$, the causality puzzle is resolved as follows. Recall that a field path integral on a trivial timefold always gives the time-ordered correlator. So when the two operators are in the asymptotic region, inside the bulk path integral each of the two terms in the commutator \eqref{eq:comm} should come with a time-ordering with respect to $g$, $\PIg{\mathcal{T}\phi(x,z)\phi(y,z)}= \PIg{\mathcal{T}\phi(y,z)\phi(x,z)}$, yielding a vanishing commutator.  Indeed, if the two points are timelike-related in the bulk, in the limit $z\to 0$ a non-time-ordered correlator would require a bulk metric with a nontrivial timefold asymptotically, violating our boundary condition. In this case, the naive bulk commutator $\PIg{[\phi(x,z),\phi(y,z)]}$ plays no role in computing the boundary commutator $\PIbdy{[\mathcal{O}(x),\mathcal{O}(y)]}$.

	Even with no timefold on the boundary, the bulk path integral includes timefolds that trivialize asymptotically~\footnote{This is related to the fact that the bulk path integral should impose gravitational constraints, necessitating integration over both positive and negative lapse \cite{Halliwell:1989dy}.}. Intuitively, the amount of time that one evolves backward goes to zero as $z\to 0$. This has no effect on the differentiate dictionary, since derivatives only act on the boundary conditions. However, for the extrapolate dictionary,  it introduces an ambiguity in the location of the bulk operators $\phi$. If there are different sheets of the bulk timefold, one has to specify which sheet the operator is on, in addition to giving its location on the sheet. If we use \eqref{eq:toextra1} one expects this ambiguity to have no effect since we take the limit $z\to 0$ for each metric where the timefold becomes trivial. 
	
	In contrast, there is a potential problem with taking $z\to0$ after integration. For any nonzero $z$, there are bulk metrics where for timelike-separated points $(x, z)$ and $(y, z)$ there is a timefold that is trivial asymptotically, but both the time-ordered and out-of-time-order correlators can be obtained by placing the operators on different sheets~\footnote{Since the bulk timefolds do not have to be minimal, a nonzero commutator can be obtained from a single timefold, by moving the earlier operator from one sheet to another. For example, on the right of Fig.\;\ref{fig:foldillus}, one can compute the commutator by moving $\phi(p)$ to the first sheet.}.
	In other words, the bulk contains a timefold that allows a nonzero commutator, but in the region outside the location of the operators, the timefold decays and completely disappears at the boundary. Under these conditions, it would appear that the integrand can be nonzero, and one again is left with the puzzle of why the integral over metrics vanishes exactly.

	However, the above occurs only if we allow bulk operators to change sheets depending on their ordering in a correlator. Since the choice of sheet is not fixed by the boundary correlator that we are trying to reproduce, one can adopt the rule that one fixes the ambiguity in the extrapolate dictionary by picking one sheet of a timefold for each operator independent of the location of the operator in the correlator. With this understanding, the integrand remains zero for each $g$ (even before taking $z\to0$), consistent with boundary causality.
	
	Finally, we extend the ordering prescription for the bulk path integrand to general $n$-point functions, allowing for timefolds inside the bulk. This upgrades \eqref{eq:toextra1} to
	\begin{equation}
		\label{eq:toextra1P}
		G \propto
		\int \mathcal{D} g \, e^{\im S_g} \lim_{z\to0} z^{\Delta_\Sigma} \PIg{\mathcal{P}\phi_n(x_n,z)\cdots\phi_1(x_1,z)}
	\end{equation}
	and analogously if $z\to0$ is taken after integration. Here $g$ is restricted to metrics that asymptote to the minimal boundary timefold required to order the field theory correlator correctly and $\mathcal{P}$ is the ordering operator enforced by the field path integral on $g$, which reduces to the time-ordering operator $\mathcal{T}$ when $g$ is a trivial timefold (cf. the order followed by the arrows in Fig.\;\ref{fig:foldillus}).

	\section{Discussion}
	In this paper, we have explained how the bulk theory manages to respect causality of a unitary boundary theory despite the bulk path integral involving boundary causality violating metrics. To do so, we formulated the bulk computation of correlators using minimal timefolds, where it becomes manifest that boundary microcausality prevails, i.e., local operators still commute at spacelike separation. 
	With non-minimal timefolds we would still expect microcausality to hold so long as the boundary theory is causal and unitary. This way, however, one would have to rely on unexpected cancellations in the integral over metrics  -- in this sense, minimal timefolds are nothing but a simpler representation of the problem (cf. charge conservation via symmetry arguments vs checking all orders in perturbation theory).
	
	Our results apply to both the differentiate dictionary as given in \eqref{eq:diffcon} and the extrapolate dictionary in the form \eqref{eq:toextra1}. It also applies to the latter if one integrates first, with a suitable rule for how to place operators on bulk timefolds that are trivial at the boundary. The basic reason for this is that a nonzero commutator for two asymptotic bulk operators requires a bulk timefold that is nontrivial at infinity, but the minimal timefold for two spacelike-separated points on the boundary is trivial. So the bulk dual of the commutator of two field theory operators at spacelike-separated points vanishes.
	
	The bulk dual of the commutator of two stress energy tensors on the boundary does not involve bulk matter fields. Nevertheless our argument using the differentiate dictionary still applies. Since the boundary metric is the source for the stress tensor, one should compute the path integral over bulk geometries with a general boundary metric. One then functionally varies the boundary metric at the location of the stress tensors. If they are spacelike-related, there is no timefold on the boundary and hence the commutator vanishes. (There is an issue with even defining the extrapolate dictionary nonperturbatively in this case arising from the apparent need to split the metric degrees of freedom into background and fluctuations.) Our argument also applies to fermionic operators which anticommute at spacelike separations. For the differentiate dictionary, the sign difference comes from the fermionic sources for these operators anticommuting.
	
	Although off-shell boundary causality violating metrics do not pose a problem to microcausality of the field theory, such metrics do contribute to general correlators. It would be interesting to understand what, if any, are the implications of this quantum gravity effect. We do not believe there are examples of holography in which a bulk classical solution violates boundary causality, but if there are, the boundary theory would have to be acausal or non-unitary. In this case, computations are timefold-dependent and our resolution does not apply.
	
	Besides microcausality, there are other properties the field theory is expected to have, which in general impose further constraints on the bulk state classically and semiclassically. For example, the invariance of von Neumann entropy under unitary transformations requires that the causal wedge be inside the entanglement wedge \cite{Wall:2012uf, Headrick:2014cta}. (This is stronger than and implies boundary causality \cite{Akers:2016ugt}.) Another requirement states that, for two causally-related bulk points, boundary regions encoding each of them cannot be totally spacelike \cite{Berenstein:2020cll}. (This extends our discussion from local operators to operators supported in subregions.) In a nonperturbative gravitational path integral there will be configurations violating this. It would be interesting to understand how the bulk path integral preserves these properties.

	
	\begin{acknowledgements}
		We would like to thank 
		Raphael Bousso,
		Ven Chandrasekaran,
		Netta Engelhardt,
		Sebastian Fischetti,
		Daniel Harlow,
		Hong Liu,
		Don Marolf
		and 
		Huajia Wang
		for discussions.
		We also thank Raphael Bousso for pointing out the role of causal diamonds in minimal timefolds, and Don Marolf for useful insights on the lapse function in bulk timefolds and for valuable comments on an earlier draft. This work was supported in part by NSF grant PHY1801805. 
	\end{acknowledgements}

	\bibliography{referencesPRL}

\end{document}